# The Structure of Graphene on Graphene/C$_{60}$/Cu Interfaces: A Molecular Dynamics Study

Alexandre F. Fonseca[1], Sócrates O. Dantas[1,2], Douglas S. Galvão[1,3], Difan Zhang[4] and Susan B. Sinnott[4,5]

[1] Applied Physics Department, Institute of Physics "Gleb Wataghin", State University of Campinas, Campinas, SP, 13083-970, Brazil.
[2] Physics Department, Science Institute, Universidade Federal de Juiz de Fora, Juiz de Fora, MG, 36036-330, Brazil.
[3] Center for Computing in Engineering and Science (CCES), State University of Campinas, Campinas-SP, 13083-959, Brazil.
[4] Department of Materials Science and Engineering, The Pennsylvania State University, University Park, PA 16801, United States.
[5] Department of Chemistry, The Pennsylvania State University, University Park, PA, 16801, United States.

E-mail: afonseca@ifi.unicamp.br

**Abstract**

Two experimental studies reported the spontaneous formation of amorphous and crystalline structures of C$_{60}$ intercalated between graphene and a substrate. They observed interesting phenomena ranging from reaction between C$_{60}$ molecules under graphene to graphene sagging between the molecules and control of strain in graphene. Motivated by these works, we performed fully atomistic reactive molecular dynamics simulations to study the formation and thermal stability of graphene wrinkles as well as graphene attachment to and detachment from the substrate when graphene is laid over a previously distributed array of C$_{60}$ molecules on a copper substrate at different values of temperature. As graphene compresses the C$_{60}$ molecules against the substrate, and graphene attachment to the substrate between C$_{60}$s ("C$_{60}$s" stands for plural of C$_{60}$) depends on the height of graphene wrinkles, configurations with both frozen and non-frozen C$_{60}$s structures were investigated in order to verify the experimental result of stable sagged graphene when the distance between C$_{60}$s is about 4 nm and height of graphene wrinkles is about 0.8 nm. Below the distance of 4 nm between C$_{60}$s, graphene becomes locally suspended and less strained. We show that this happens when C$_{60}$s are allowed to deform under the compressive action of graphene. If we keep the C$_{60}$s frozen, spontaneous "blanketing" of graphene happens only when the distance between them are equal or above 7 nm. Both above results for the existence of stable sagged graphene for C$_{60}$ distances of 4 or 7 nm are shown to agree with a mechanical model relating the rigidity of graphene to the energy of graphene-substrate adhesion. Although the studies of intercalation of molecules on interfaces formed by graphene-substrate are motivated by finding out ways to control wrinkling and strain in graphene, our work reveals the shape and structure of intercalated molecules and the role of stability and wrinkling on final structure of graphene. In particular, this study might help the development of 2D confined nanoreactors that are considered in literature to be the next advanced step on chemical reactions.

## 1. Introduction

Recently, a new strategy to create and control local strains on graphene has been developed by Monazami *et al*. [1]. It is based on intercalating $C_{60}$ molecules between graphene and a substrate. It was demonstrated that the concentration of $C_{60}$ molecules and temperature are the main variables to control local strain in graphene. They also observed the formation of crystalline and amorphous patterns of $C_{60}$s ("$C_{60}$s" stands for plural of $C_{60}$) under graphene, and that blanketing of graphene around $C_{60}$s on the substrate occurs whether the distance between the molecules was, at least, 4 nm. This is one example of a known mechanism to modify properties of layered materials based on intercalation of molecules, which is well known for graphite and it is called Graphite Intercalation Compounds (GIC) [2]. One GIC example is the superconducting properties obtained by intercalating $C_{60}$s and alkali metals into graphite [3]. As highlighted by Daukiya *et al*. [4], this mechanism is being used on systems formed by graphene-metal [1][5][6][7][8] and few layers of graphene [9][10][11], and it is considered to be one of the newest forms to promote chemical reactions and catalysis at nanoscale [11][12].

In another set of experiments, Lu *et al*. [7] initially placed $C_{60}$ molecules on a ruthenium substrate and, after that, they grew graphene on top of them. They found out that at high temperatures (about 1000 K or more), some $C_{60}$s reacted with each other under graphene, and formed another segment of graphene sheet or a bilayered graphene structure. They also observed the coexistence of bilayered graphene and layers of graphene on remaining $C_{60}$ molecules, including the blanketing of graphene around $C_{60}$s on the substrate, if the $C_{60}$s distance is larger than ~ 4.5 nm.

Motivated by these two experiments [1][7], here we investigated the formation of the "blanketing" of graphene on a set of $C_{60}$ molecules already placed on a copper substrate, using tools of classical molecular dynamics (MD) simulations. The above experiments cannot fully reveal the local atomistic features underneath the "blanketing" process. We also have a limitation. We cannot simulate the process of the $C_{60}$ intercalation because the diffusion time of $C_{60}$s to get in between graphene and copper substrate is much longer than what is feasible by MD simulation methods. However, we can simulate the dynamics of structures similar to what Lu *et al*. [7] studied or those obtained by Monazami *et al*. [1] *after* the intercalation of $C_{60}$s, i.e., we can start from a configuration where the graphene is placed on top of an array of $C_{60}$s that is already placed on top of a substrate to investigate the formation of wrinkles along the graphene structure.

Monazami *et al*. [1] reported that they measured different values of heights of the $C_{60}$ molecules under graphene. This observation lead us to test the "blanketing" of graphene on two system configurations *with* and *without* freezing the array of $C_{60}$ molecules. Then, we verified that wrinkled graphene sagged and adhered to the substrate in the space between the $C_{60}$s when the $C_{60}$-to-$C_{60}$ distances, *d*, are, at least, 7 nm, if the $C_{60}$s are frozen, and 4 nm if they are not frozen. The reason for that will be shown to be related to the height of the molecules, which can change due to the graphene compression against the substrate. Fully spontaneous "blanketing" of graphene was observed for frozen $C_{60}$s at distances $d \cong 7$ nm. For frozen $C_{60}$s and distances $d < 7$nm, we investigated the dependence of the spontaneous formation of "blanketing" of graphene with the temperature, and we found out that decreasing the distance, *d*, larger temperature values are needed to promote even partial blanketing of graphene, as observed in the experiments [1][7].

In the next sections, we describe the computational details, present the results and summarize the main conclusions.

## 2. Computational Details

We carried out fully atomistic MD simulations using the third generation of the Charge Optimized Many Body (COMB3) potential [13][14] to investigate systems in which graphene covers a set of $C_{60}$ molecules on a (100) surface of a copper substrate. The COMB3 potential used in this work is parametrized to address both hydrocarbon and copper and copper-oxide structures [15][16]. It is able to simulate multicomponent systems and has been successfully used to study similar systems to those of the present work as, for example, the formation of graphene wrinkles on copper substrates due to carbon-copper different thermal expansion coefficients [17], and the structure and thermal stability of graphene-titanium interfaces on top of copper and copper-oxide substrates [18][19]. The computational package LAMMPS [20][21], which includes the COMB3 potential, was used to perform the numerical integration of the equations of motion. The advantage of using COMB3 is that it has full parametrization for carbon-carbon (C-C), copper-copper (Cu-Cu) and copper-carbon (Cu-C) interactions, without the need to use different potentials for each part of the system.

The systems studied here are illustrated by the model shown in Figure **1**. It consists of a large copper substrate, a set of sixteen $C_{60}$ molecules initially placed in a square array of distance, *d*, and a graphene sheet on top of them. Periodic boundary conditions (PBC) were used to the planar dimensions of the copper substrate. Systems with different arrays of $C_{60}$ molecules have different sizes of copper substrate and graphene sheets. The sets of square arrays of $C_{60}$s studied here are for distances varying from 4 to 7 nm.

MD simulations were performed with a Langevin thermostat [22] and damping factor fixed in 100 fs, and with the timestep set in 0.2 fs. All structures were initially obtained by energy minimizations with force and relative energy tolerance of $10^{-12}$ eV/Å and $10^{-14}$, respectively. In all MD simulations, the bottom two layers of copper substrate were kept fixed to mimic the bulk system.

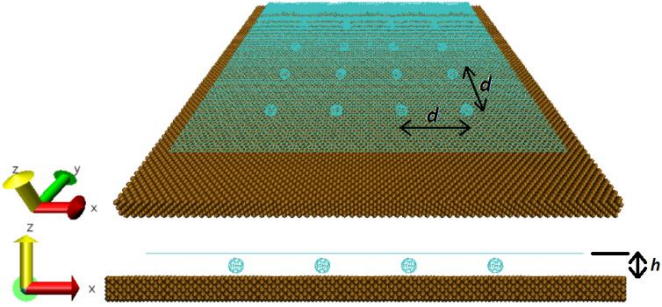

**Figure 1**: Two views (in perspective on top and lateral at the bottom) of the system model formed by a copper substrate cleaved along the (100) crystallographic direction, a square array of $C_{60}$s molecules and a graphene sheet initially on top of the system. The sizes of the substrate, graphene and distance, *d*, between $C_{60}$s vary with the system studied. *h* is the distance from graphene to the copper substrate and will be the distance of the graphene wrinkles when it occurs. Copper and carbon atoms are drawn in brown van der Waals and cyan line colours, respectively. Some graphene sheets were passivated by hydrogen (not shown above) for comparison.

We performed two tests. One test is on the stability of the graphene wrinkles that sag and touch the copper substrate between $C_{60}$ molecules in a square array shown in Figure **1** with $d \cong 4$ nm. The stability of graphene wrinkles can be studied by the mechanical model developed by Yamamoto *et al*. [23]; for the equilibrium distances, $d_{EQ}$, between wrinkles having deflection heights, *h*, as given by:

$$d_{EQ} = h \left(\frac{E_{2D}}{G}\right)^{\frac{1}{4}} \cong 5.04\, h, \quad (1)$$

where $E_{2D}$ is the tensile rigidity ($2.12 \times 10^3$ eV/nm$^2$ [24]) and *G* is the graphene-copper adhesion energy per area (3.28 eV/nm$^2$ [17]), and the second equality comes from substituting $E_{2D}$ and *G* in the equation. *h* will be the distance between graphene and substrate and it can be written as

$$h = f + 0.34, \quad (2)$$

where *f* is the diameter of the $C_{60}$ and the 0.34 factor (in nm) is the approximate value for the distance between $C_{60}$ and copper plus distance of $C_{60}$ to graphene. The prediction from the model given by Eq. (1) is that stable graphene wrinkles around $C_{60}$ molecules with graphene remaining sagged to the substrate occur for distances between $C_{60}$s, $d_{EQ}$, of 5 (4) nm for wrinkle heights, *h*, of 1 (0.8) nm.

$d \cong 4$ (4.5) nm is the smallest distance between $C_{60}$ molecules on copper (ruthenium) substrate observed by Monazami *et al*. [1] (Lu *et al*. [7]) for which graphene presented stable wrinkles and sagged parts between $C_{60}$s. Here, we evaluate the relationship between the height of the graphene wrinkles and their stability, by simulating the whole system in two situations: *with* and *without* freezing the $C_{60}$ atomic structure, *i.e.,* *with* and *without* freezing the height of the wrinkles. Monazami *et al*. [1] found out values of heights of the graphene wrinkles that are smaller than the $C_{60}$ diameter when it is isolated. As graphene sheet compresses the $C_{60}$ molecules against the substrate, their structure might become smashed and, therefore, the graphene wrinkle height decreased. In this test, we manually deformed the graphene sheet by moving down the regions between the $C_{60}$ positions, or "sagging" them until they touch the substrate, while the rest of the system is kept fixed. Then, these regions were kept fixed and the rest of the system was allowed to equilibrate by the application of an energy minimization method, followed by MD simulations at temperature values of 1, 10 and 100 K by about 20 ps each. Until this point, the $C_{60}$s and the moved regions of graphene were kept frozen. After that, we performed another set of MD simulations of the whole system, without the constraints on the initially moved graphene regions, at 1, 300 and 600 K for about 20 ps each, in order to verify the stability of the sagged graphene wrinkles in both situations: with or without freezing the $C_{60}$s.

An additional test investigated the spontaneous formation and stability of the "blanketing" process of graphene sheet around the $C_{60}$s on a fixed square array with different distances, *d*. For this case, the borders of the graphene sheets are hydrogen passivated and every structure was MD simulated for a total time between 2 and 4 ns, depending on the time required for the graphene sheet to "blanket" the system. The systems were simulated at 300, 700 and 1000 K. Lu *et al*. [7] investigated the reactions occurring to the $C_{60}$s under the compression of graphene sheet and thermal fluctuations. Here, we are not going to investigate the C60s reactions but only the $C_{60}$-to-$C_{60}$ distances that allow graphene to spontaneous blanket the $C_{60}$ molecules and its relative stability, at different temperatures, of the system with graphene initially placed on top of the array of $C_{60}$s. The Yamamoto *et al*. [23] model will be used to interpret the results and analyses of the experiments, in terms of the equilibrium distances between wrinkles in graphene.

These results will be presented and discussed in the next section.

## 3. Results and Discussion

In Figure **2** we present the results for the first test where $d \cong 4$ nm (see Figure **1** for the meaning of *d*). When the $C_{60}$ molecules were frozen, the graphene sheet that was initially and manually deformed to blanket each $C_{60}$ on the substrate, simply detached from it forming the two plateau-like structure showed on the top panel of Figure **2**. However, when the $C_{60}$s in this test are not frozen, they get smashed by graphene sheet compression and the wrinkles and sags remain, as we can see on the bottom panel of Figure **2**. We estimated the value of the final $C_{60}$ height as about 0.4 nm, when it is not frozen.

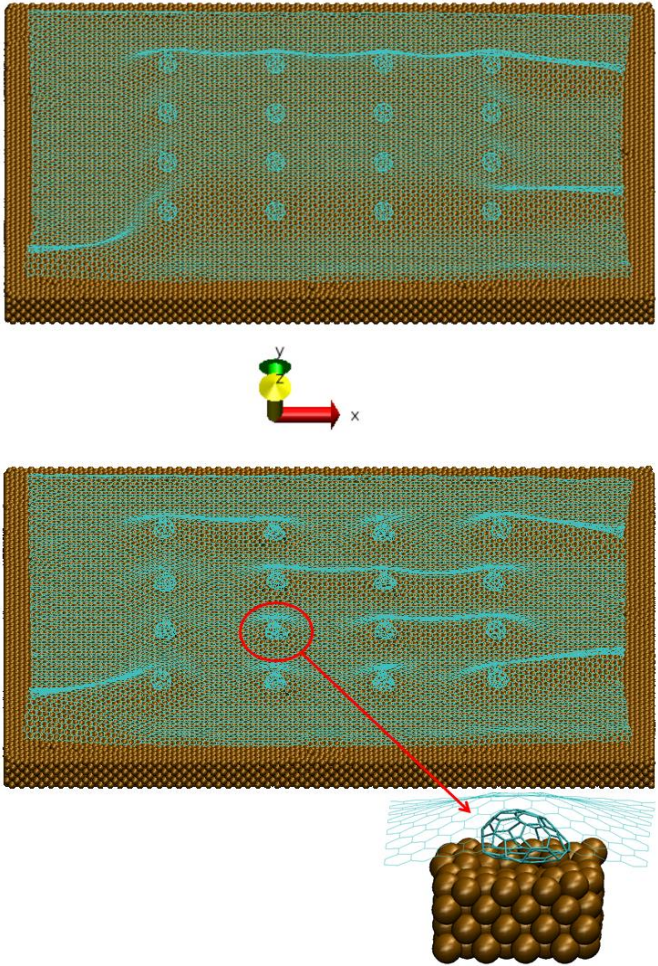

**Figure 2**: MD snapshots of equilibrated structures of a graphene sheet of 21 x 23 nm of size on a set of 16 frozen (top) and non-frozen (bottom) $C_{60}$ molecules in a square array of $d$ = 4 nm, on a copper (100) substrate of 23 x 25 nm. On top, the graphene sheet formed a plateau-like structure covering all $C_{60}$ molecules. At the bottom, the graphene sheet remained blanketing the $C_{60}$ molecules on the substrate after 20 ps of simulation at 1 K (and even after simulations at 300 and 600 K, data not shown). The structure inside the red circle is zoomed in order to provide a better view of the local atomic structure of a smashed $C_{60}$ molecule induced by graphene compression. Copper and carbon atoms are shown in brown/van der Waals spheres and cyan/lines (graphene) or cyan/sticks ($C_{60}$), respectively.

The height, $h$, of graphene wrinkle can be estimated by eq. (2). If the $C_{60}$ is (not) kept frozen, $h$ = 1 + 0.34 = 1.34 nm ($h$ = 0.4 + 0.34 = 0.74 nm). The model proposed by Yamamoto *et al*. [23] predicts the stability of blanketed graphene on a square array of $C_{60}$s with $d_{EQ}$ = 4 nm only if the wrinkles height were ≤ 0.8 nm. Top and bottom panels of Figure **2** show that blanketing of graphene remained stable only for the non-frozen $C_{60}$s structure, where the above condition (height ≤ 0.8 nm and $d_{EQ}$ = 4 nm) is satisfied, as expected. We further increase the temperature of the system shown on the bottom panel of Figure **2** to 300 and, after that, to 600 K, in order to verify its thermal stability (data not shown). This is also consistent to Monazami *et al*. results (see figure 4(a) of [1]), where they measured apparent wrinkle heights between 0.6 and 0.8 nm for distances between $C_{60}$s of about 4 nm. Therefore, the deformation of $C_{60}$s under the graphene compression is consistent with the experiments.

The results for the spontaneous formation of the blanketed graphene around frozen $C_{60}$s on a square array of $C_{60}$-to-$C_{60}$ distances varying from 5 to 7 nm are presented now. For $d$ = 5 nm, we observed that even at the highest temperature value considered here, $T$ = 1000 K, only the plateau-like structure is formed, similar to the top panel structure shown in Figure **2**.

The structures having $d$ = 6 nm and $d$ = 7 nm were simulated for several nanoseconds at temperatures of 300, 700 and 1000 K, respectively. For $d$ = 6 nm, we have observed no blanketing formation of graphene around the $C_{60}$s for $T$ = 300 K (data not shown) and for $T$ = 700 K (structure shown on the top panel of Figure **3**). However, for $T$ = 1000 K, we have found spontaneous partial blanketing formation. In the bottom panel of Figure **3**, the structure for $d$ = 6 nm after 2 ns of simulations at 1000 K shows two regions: i) plateau-like, and; ii) graphene blanketing formation (inside the yellow circle). It shows that the $C_{60}$-to-$C_{60}$ distance is not yet in the minimum threshold to allow the blanketing formation of graphene around the $C_{60}$s.

For the structures with square arrays of frozen $C_{60}$s with $d$ = 7 nm, we observed the practically full blanketing formation of graphene around the $C_{60}$s at all temperatures studied, $T$ = 300, 700 and 1000 K, and for a period of time of about < 1 ns. Top and bottom panels of figure **4** show the structures after about 1 ns of simulation at 300 and 1000 K, respectively. It shows the stability of graphene sagging on copper substrate around the $C_{60}$ molecules.

The results show that at $d$ = 7 nm, the conditions required by the Yamamoto *et al*. [23] model for the stability of graphene wrinkles are satisfied. How to interpret these results of partial (full) blanketing formation at high (or low) temperature for $C_{60}$ square arrays with $d$ = 6 (7) nm? We can also use the Yamamoto *et al*. [23] model to analyse them while taking into account our suggested definition of the wrinkle height given by eq. (2). If we use the eq. (1) to estimate the equilibrium distance between graphene wrinkles, when $h$ = 1 + 0.34 = 1.34 nm, we found $d_{EQ} \cong 6.75$ nm. So $d$ = 7 nm is above this minimum distance. However, for the structures having $d$ = 5 or 6 nm, that are smaller than the above value of 6.75 nm, none or only partial blanketing occurs at high temperature. These results are, then, consistent with the Yamamoto *et al.* [23] model as long as the eq. (2) is considered in the determination of graphene wrinkle height.

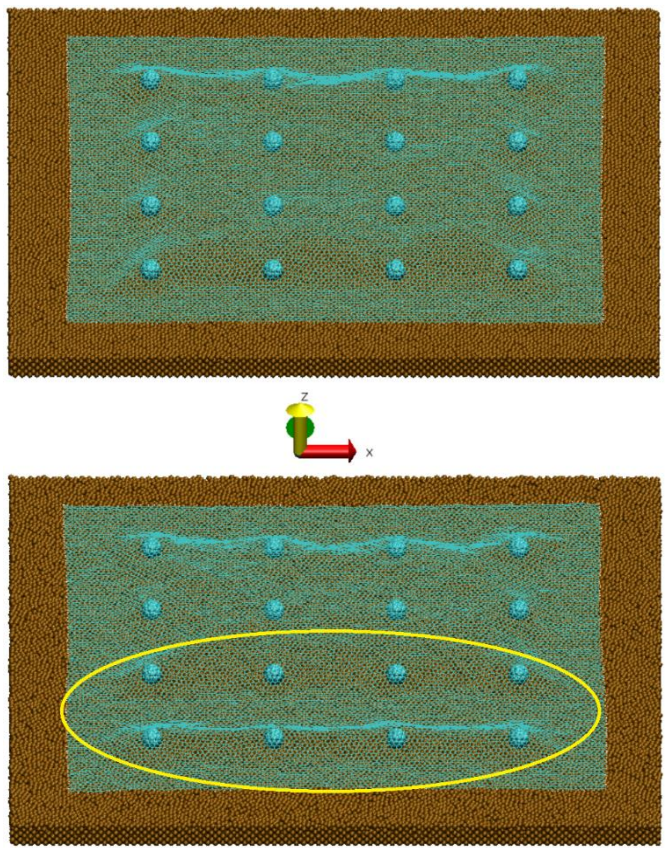

**Figure 3**: MD snapshots of equilibrated structures of a graphene sheet of 25 x 28 nm of size on a set of 16 frozen $C_{60}$ molecules in a square array of $d = 6$ nm, on a copper (100) substrate of 31 x 34 nm. Top: structure simulated at 700 K after 3 ns. Bottom: structure simulated at 1000 K after 2 ns. Yellow circle highlights the region where partial blanketing of graphene around $C_{60}$s was formed. Copper and carbon atoms of $C_{60}$ molecules are shown in brown and cyan van der Waals spheres, respectively, and graphene in cyan lines. Graphene borders are passivated by hydrogen atoms (small white lines).

## 4. Conclusions

This study presents the results of MD simulations of structures formed by graphene deposited on top of square arrays of $C_{60}$ molecules already deposited on copper substrate. We have simulated square arrays of $C_{60}$-to-$C_{60}$ distances from 4 to 7 nm, for $C_{60}$ molecules frozen and non-frozen, and at different values of temperature ranging from 1 to 1000 K. These simulations were performed based on two experiments involving the intercalation of $C_{60}$ molecules between graphene and copper substrates [1][7] and the heating of arrays of $C_{60}$ molecules on bottom of a graphene sheet at high temperatures ($T \geq 1000$ K). As the timeframe of intercalation of $C_{60}$ molecules between graphene and substrate is much larger than what can be simulated by all-atom molecular dynamics methods, we modeled the stability of the blanketed graphene around $C_{60}$ molecules for the conditions mentioned at the beginning of this paragraph.

Although the experiments take place under different conditions and started from different structures, our simulations revealed some interesting features related to those experiments [1][7] that might be useful to carry out new experiments and to better understand these processes at atomic level. The results are summarized as follows.

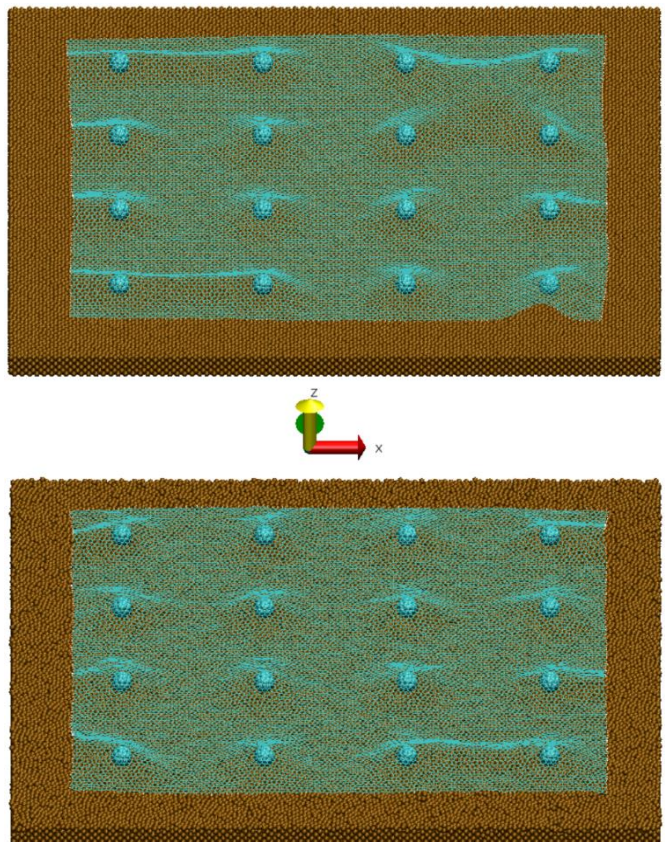

**Figure 4**: MD snapshots of equilibrated structures of a graphene sheet of 25 x 28 nm of size on a set of 16 frozen $C_{60}$ molecules in a square array of $d = 7$ nm, on a copper (100) substrate of 31 x 34 nm. Top: structure simulated at 300 K after 1 ns. Bottom: structure simulated at 1000 K after 1.3 ns. Copper and carbon atoms of $C_{60}$ molecules are shown in brown and cyan van der Waals spheres, respectively, and graphene in cyan lines. Graphene borders are passivated by hydrogen atoms (small white lines).

The experimentally observed minimum values of $C_{60}$-to-$C_{60}$ distances at which graphene remains sagged to the substrate and has its wrinkles stable, $d = 4$ (4.5) nm according to Monazami *et al.* [1] (Lu *et al.* [7]), are such that the $C_{60}$ molecules have to be smashed by graphene compression against the substrate. Otherwise, the corresponding graphene wrinkle heights would not allow for such a small equilibrium distance between the wrinkles, according to the mechanical model developed by Yamamoto *et al.* [23].

The spontaneous formation of graphene blanketing around $C_{60}$s was also investigated in order to determine the stability of the blanketed structure and, on another hand, the formation and stability of a plateau-like structure as seen in the top panels of figures **2** and **3**. We concluded that graphene on top of a structure having the right amount of spacing between the $C_{60}$ molecules, will get spontaneously blanketed around the $C_{60}$s, even for relatively low temperatures. Alternatively, when the distance between $C_{60}$s are small, we can conclude that even at moderate temperatures ranging from room to values like 700 K as considered here, it is possible to produce a confined region (the region inside the plateau-like structure of top panels of figures **2** and **3**) at nanoscale whose size can be kept constant in order to have several molecules confined, for example, to react in an isolated ambient from the external side and under some level of graphene compression.

Our results are in agreement with the similar work of Tang *et al*. [25], who performed a systematic study of the interfacial adhesion of graphene on a pillared surface. Their results illustrated the dependence of the blanketing of graphene around pillars on a substrate with the pillar-to-pillar distance and height. One substantial difference of this work relative to the work of Tang *et al*. is that we did not have to use different potentials to describe carbon-carbon, cooper-cooper and carbon-cooper interactions. COMB3 is fully parameterized to simulate those interactions, including van der Walls and Coulombic contributions.

Instead of a square array of $C_{60}$ molecules, if we have sets of carbon nanotubes on the substrate, a graphene sheet on top of this set would be expected to blanket and form sags within the free space between the tubes and also allow for the formation of regions for nanoreactions. The kind of structures studied here opens up new possibilities to investigate different forms to obtain isolated nanoregions and we hope this can motivate new research along these lines.

**Acknowledgements**

A.F.F. (grant number #311587/2018-6), D.S.G. and S.O.D. acknowledge support from the Brazilian Agencies CNPq and CAPES. A.F.F. acknowledges grant #2018/02992-4 from São Paulo Research Foundation (FAPESP) and from FAEPEX/UNICAMP. D.S.G. and S.O. D. thank the Center for Computational Engineering and Sciences at Unicamp for financial support through the FAPESP/CEPID Grant #2013/08293-7. This research also used the computing resources and assistance of the John David Rogers Computing Center (CCJDR) in the Institute of Physics "Gleb Wataghin", University of Campinas. D.Z. and S.B.S. were supported by UNCAGE-ME, an Energy Frontier Research Center funded by the U.S. Department of Energy, Office of Science, Basic Energy Sciences under Award #DE-SC0012577.

**References**

[1] Monazami E, Bignardi L, Rudolf P and Reinke P 2015 *Nano Lett*. **15** 7421. DOI: 10.1021/acs.nanolett.5b02851

[2] Chung D D L 2001 Graphite Intercalation Compounds *Encyclopedia of Materials: Science and Technology* ed K Jurgen Buschow *et al*. (London, Elsevier, 2001) p 3641.

[3] Fuhrer M S, Hou J G, Xiang X -D and Zettl A 1994 *Solid State Communications* **90** 357. DOI: 10.1016/0038-1098(94)90798-6

[4] Daukiya L, Nair M N, Cranney M, Vonau F, Hajjar-Garreau S, Aubel D and Simon L 2018 *Progress in Surface Science* in press. DOI: 10.1016/j.progsurf.2018.07.001

[5] Cranney M, Vonau F, Pillai P B, Denys E, Aubel D, De Souza M M, Bena C and Simon L 2010 *EPL* **91** 66004. DOI: 10.1209/0295-5075/91/66004

[6] Varykhalov A, Gudat W and Rader O 2010 *Adv. Mater*. **22** 3307. DOI: 10.1002/adma.201000695

[7] Lu J, Zheng Y, Sorkin A and Loh K P 2012 *Small* **8** 3728. DOI: 10.1002/smll.201201113

[8] Erturk A S, Yildiz Y O and Kirca M 2018 *Computational Materials Science* **144** 193. DOI: 10.1016/j.commatsci.2017.12.033

[9] Kirca M 2015 *Composites Part B* **79** 513. DOI: 10.1016/j.compositesb.2015.04.050

[10] Verhagen T G A, Vales V, Kalbac M and Vejpravova J 2017 *Diamond & Related Materials* **75** 140. DOI: 10.1016/j.diamond.2017.03.001

[11] Mirzayev R, Mustonen K, Monazam M R A, Mittelberger A, Pennycook T J, Mangler C, Susi T, Kotakoski J and Meyer J C 2017 *Sci. Adv*. **3** e1700176. DOI: 10.1126/sciadv.1700176

[12] Fu Q and Bao X 2017 *Chem. Soc. Rev*. **46** 1842. DOI: 10.1039/c6cs00424e

[13] Liang T, Shan T –R, Cheng Y –T, Devine B D, Noordhoek M, Li Y, Lu Z, Phillpot S R, Sinnott S B 2013 *Mat. Sci. and Eng. R* **74** 255. DOI: 10.1016/j.mser.2013.07.001

[14] COMB3. Available at https://research.matse.psu.edu/sinnott/software (accessed 29 January 2019).

[15] Devine B, Shan T –R, Cheng Y –T, McGaughey A J H, Lee M, Phillpot S R and Sinnott S B 2011 *Phys. Rev. B* **84** 125308. DOI: 10.1103/PhysRevB.84.125308

[16] Liang T, Cheng Y –T, Nie X, Luo W, Asthagiri A, Janik M J, Andrews E, Flake J and Sinnott S B 2014 *Catal. Commun*. **52** 84. DOI: 10.1016/j.catcom.2013.11.033

[17] Klaver T P C, Zhu S –E, Sluiter M H F and Janssen G C A M 2015 *Carbon* **82** 538. DOI: 10.1016/j.carbon.2014.11.005

[18] Fonseca A F, Liang T, Zhang D, Choudhary K, Phillpot S R and Sinnott S B 2017 *ACS Appl. Mater. Interfaces* **9**, 33288. DOI: 10.1021/acsami.7b09469

[19] Fonseca A F, Liang T, Zhang D, Choudhary K, Phillpot S R and Sinnott S B 2018 *MRS Advances* **3** 457. DOI: 10.1557/adv.2018.160

[20] Plimpton S 1995 *J. Comput. Phys*. **117** 1. DOI: 10.1006/jcph.1995.1039

[21] LAMMPS - Molecular Dynamics Simulator. Available at http://lammps.sandia.gov (accessed 29 January 2019).

[22] Schneider T and Stoll E 1978 *Phys. Rev. B* **17** 1302. DOI: 10.1103/PhysRevB.17.1302

[23] Yamamoto M, Pierre-Louis O, Huang J, Fuhrer M S, Einstein T L and Cullen T L 2012 *Phys. Rev. X* **2** 041018. DOI: 10.1103/PhysRevX.2.041018

[24] Lee C, Wei X, Kysar J W and Hone J 2008 *Science* **321** 385. DOI: 10.1126/science.1157996
[25] Tang X, Zhang K, Deng X, Zhang P and Pei Y 2016 *Molecular Simulation* **42** 405. DOI: 10.1080/08927022.2015.1059430